\documentclass[aps,prb,twocolumn]{revtex4}
\usepackage{graphicx,epsf}
\usepackage{amsmath}

\newcommand{\ybco}{YBa$_2$Cu$_3$O$_{6+\delta}$}

\newcommand{\ybcoc}{YBa$_2$Cu$_3$O$_{6.45}$}

\newcommand{\bscco}{Bi$_2$Sr$_2$CaCu$_2$O$_{8+\delta}$}

\newcommand{\ccoc}{Ca$_{2-x}$Na$_x$CuO$_2$Cl$_2$}

\newcommand{\srruo}{Sr$_3$Ru$_2$O$_7$}

\newcommand{\e}{\epsilon}

\begin{document}

\title{
Nernst effect anisotropy as a sensitive probe of Fermi surface distortions\\
from electron-nematic order
}

\author{Andreas Hackl}
\author{Matthias Vojta}
\affiliation{
Institut f\"ur Theoretische Physik, Universit\"at zu K\"oln,
Z\"ulpicher Stra\ss e 77, 50937 K\"oln, Germany
}
\date{November 24, 2009}

\begin{abstract}
We analyze the thermoelectric response in layered metals with spontaneously broken rotation
symmetry. We identify the anisotropy of the quasiparticle Nernst signal as an extremely
sensitive probe of Fermi surface distortions characteristic of the ordered state.
This is due to a subtle interplay of different transport anisotropies which become
additionally enhanced near van-Hove singularities.
Applied to recent experiments, our results reinforce the proposal that the underdoped cuprate
superconductor \ybco\ displays such ``electron-nematic'' order in the pseudogap regime.
\end{abstract}
\pacs{72.15.Jf,71.10.Hf,74.72.-h,74.25.Fy}

\maketitle


Spontaneous breaking of lattice rotation symmetry due to electronic correlations is
currently in the focus of intense interest, most prominently in
cuprate high-temperature superconductors such as
\ybco\ (Refs.~\onlinecite{hinkov08a,taill10a,taill10b})
and in the metamagnetic metal \srruo\ (Ref.~\onlinecite{borzi07}).
In analogy to liquid crystals, a phase with broken rotation (but preserved translation)
symmetry has been dubbed electron nematic.\cite{KFE98}

In cuprates, electron-nematic order has been discussed early on as intermediate phase which
occurs upon melting of a uni-directional charge-density-wave (``stripe'')
phase.\cite{KFE98,kiv_rmp,mv_ap}
Microscopically, it is one of the known instabilities of the two-dimensional (2d) Hubbard
model.\cite{YK00,HM00,yamase06}
The first clear-cut signature of electron-nematic order in cuprates was found in
neutron scattering experiments on \ybcoc,\cite{hinkov08a}
where the spin-fluctuation spectrum was found to develop a distinct anisotropic
incommensurability below a temperature $T$ of about 150\,K.
Earlier transport measurements on \ybco\ (Ref.~\onlinecite{ando02}) also detected resistivity
anisotropies $\rho_a/\rho_b$ of order 2 in the underdoped regime,
but remained less conclusive.
The reason is that the crystal structure of \ybco\ contains CuO chains which break the
otherwise tetragonal symmetry of the CuO$_2$ planes.
In an order-parameter language, this implies a small field which couples linearly to
the nematic order parameter.
This has two main effects:
(i) potentially existing nematic order will be aligned and
(ii) a nematic ordering transition will be smeared out.
While (i) enables observables to show a macroscopic
anisotropy, which might otherwise be masked by domain formation,
(ii) implies that {\em electronic} nematic order and purely structural effects
cannot be sharply distinguished.
Remarkably, locally broken rotation symmetry has been found on the surface of \bscco\ and \ccoc\
using scanning tunneling microscopy.\cite{kohsaka07}
In \srruo, nematic order is a candidate explanation for the low-$T$ phase
which masks the metamagnetic critical endpoint at around 8\,T.\cite{grigera04,borzi07}
Resistivity anisotropies have been detected here, but a full picture has not yet emerged,
because the rather unusual thermodynamics near the low-$T$ phase boundaries
is not understood.

Very recent measurements\cite{taill10a,taill10b} of the Nernst effect,\cite{nernst_fl}
that is the transverse voltage induced by a thermal gradient in the presence of a magnetic field,
have uncovered a surprisingly large anisotropy in \ybco\ below the
pseudogap temperature.
Consequently, this has been interpreted as evidence for broken rotation symmetry,
but a theory which links this anisotropy to a particular ordered state was not
available.
Previously, the Nernst effect had been extensively studied as a probe of cuprate pseudogap
physics,\cite{wang06} but mainly interpreted as sign of precursor superconducting
fluctuations.
Only recently, a second contribution to the Nernst signal has been identified and
associated to quasiparticle physics.\cite{taill09}
Theoretically, the quasiparticle Nernst effect in the presence of
antiferromagnetic\cite{hackl09a} and stripe\cite{hackl09b} order has been investigated.

In this paper, we analyze, for the first time, the spatial anisotropy of the Nernst
effect. We show that an enhanced and strongly anisotropic quasiparticle Nernst signal
arises from the Fermi surface distortions which accompany broken rotation symmetry in metals.
Surprisingly, small anisotropies in the kinetic energy can be easily enhanced by an order
of magnitude in the Nernst signal, due the interplay of the anisotropies of conductivity
and thermopower.
A similarly large anisotropy does {\em not} arise from stripe phases.
Hence, we establish the Nernst effect as a very sensitive and unique probe of electron-nematic
order.\cite{directfoot}
Applied to the measurements on underdoped \ybco\ (Refs.~\onlinecite{taill10a,taill10b}),
we confirm the presence of the long-sought electron-nematic phase in the pseudogap regime.


{\it Quasiparticle model.}
We shall calculate the thermoelectric response in a one-band model of quasiparticles.
For concreteness, we work on a two-dimensional (2d) square lattice with a tight-binding dispersion
\begin{eqnarray}
\varepsilon_{\bf k} = &-& 2 t_1 (\cos k_x +\cos k_y ) - 4 t_2 \cos k_x \cos k_y \nonumber\\
                &-&2 t_3 (\cos 2 k_x +\cos 2 k_y) - \mu\,.
\label{dispersion}
\end{eqnarray}
Such a 2d model is appropriate not only for cuprate superconductors, but (with some
modifications) for a variety of other layered materials with tetragonal lattice structure,
including \srruo. To simplify calculations, we shall neglect inter-layer coupling, but
will comment on its effect towards the end.

Nematic order distorts the band structure, which, on the mean-field level,
can be captured by anisotropic hopping parameters.\cite{YK00,yamase06}
Here we shall focus on the case of a $d_{x^2-y^2}$ electron nematic, as may arise from a
Pomeranchuk instability in the $l\!=\!2$ channel or as precursor of a stripe phase.
We introduce an anisotropy parameter $\e$, such that the hopping matrix elements obey
$t_{1x,y} = (1\pm\e/2) t_1$ and $t_{3x,y} = (1\pm\e/2) t_3$.
For $\e\neq 0$, the lattice symmetry is thus broken from $C_4$ down to $C_2$.


{\it Transport theory.}
The linear thermoelectric response is captured by three conductivity tensors
$\hat{\sigma}$, $\hat{\alpha}$, and $\hat{\kappa}$, which relate charge current $\vec{J}$
and heat current $\vec{Q}$ to electric field, $\vec{E}$ and thermal gradient,
$\vec{\nabla }T$ vectors:
\begin{equation}
\left(
\begin{array}{c}
\vec{J} \\
\vec{Q} \\
\end{array}
\right)
=
\left(
\begin{array}{cc}
\hat{\sigma}  & \hat{\alpha} \\
T\hat{\alpha} &  \hat{\kappa}\\
\end{array}
\right)
\left(
\begin{array}{c}
\vec{E} \\
-\vec{\nabla} T\\
\end{array}
\right).
\label{thermoelectrics}
\end{equation}
The electrical field induced by a thermal gradient in the absence of an
electrical current can be expressed by the linear response relation
$\vec{E}= -\hat{\vartheta } \vec\nabla T$, and
Eq. \eqref{thermoelectrics} together with $\vec{J}=0$
yields $\vec{E}=\hat{\sigma}^{-1} \hat{\alpha} \vec\nabla T$.
Therefore, the Nernst signal $\vartheta_{yx}$,
defined as the {\em transverse} voltage $E_y$ generated by a thermal gradient $\nabla_x T$,
reads
\begin{equation}
\vartheta_{yx}=-\frac{\sigma_{xx}\alpha_{yx}-\sigma_{yx}\alpha_{xx}}{\sigma_{xx}\sigma_{yy}-\sigma_{xy}\sigma_{yx}}
\label{nernstsignal}
\end{equation}
and $\vartheta_{xy}$ is obtained from $x\leftrightarrow y$.
For a magnetic field $\vec B = B\hat z$ in $z$ direction,
the Nernst { \it coefficient \/} is usually defined as $\nu_{yx}=\vartheta_{yx}/B$,
which tends to become field-independent at small $B$.
We employ a sign convention such that the vortex Nernst coefficient is always positive
(formally $\nu_{xy}=-\vartheta_{xy}/B$, Ref.~\onlinecite{signnote}).
In general, the Nernst coefficient can be negative or positive, for example if it is caused by the
flow of charged quasiparticles.

The simplest description of quasiparticle transport is via the Boltzmann equation
in the relaxation-time approximation. We assume a momentum and energy-independent
relaxation time $\tau_0$, as is appropriate for elastic impurity scattering at low
temperatures.
Then, the electrical and thermoelectrical conductivities read:
\begin{eqnarray}
\alpha_{xx} &=&  \frac{2e\tau_0}{T} \sum_{{\bf k}} \frac{\partial f_{\bf k}^0}{\partial \varepsilon({\bf k})}
\varepsilon({\bf k})  (v_{\bf k}^x)^2,    \nonumber\\
\alpha_{xy} &=&  \frac{2e^2B\tau_0^2}{T\hbar c} \sum_{{\bf k}} \frac{\partial f_{\bf k}^0}{\partial \varepsilon({\bf k})}
\varepsilon({\bf k}) v_{\bf k}^x
\biggl[ v_{\bf k}^y \frac{\partial v_{\bf k}^y}{\partial k_x} -  v_{\bf k}^x \frac{\partial v_{\bf k}^y}{\partial k_y}\biggr],\nonumber\\
\sigma_{xx} &=&  -2e^2\tau_0 \sum_{{\bf k}} \frac{\partial f_{\bf k}^0}{\partial \varepsilon({\bf k})} (v_{\bf k}^x)^2 , \nonumber\\
\sigma_{xy} &=& -2 \frac{e^3 B\tau_0^2}{\hbar c} \sum_{{\bf k}} \frac{\partial f_{\bf k}^0}{\partial \varepsilon({\bf k})}
v_{\bf k}^x  \biggl[ v_{\bf k}^y \frac{\partial v_{\bf k}^y}{\partial k_x} -  v_{\bf k}^x \frac{\partial v_{\bf k}^y}{\partial k_y}\biggr],
\label{boltzmann}
\end{eqnarray}
where $e=|e|$ is the electron charge, ${\bf v}_{\bf k}$ the velocity, and $f_{\bf k}^0$
the Fermi function.
In this approximation, the Nernst signal is linear in both $T$ (at low $T$) and $\tau_0$.
In the following, we shall be interested in the magnitude and anisotropy of the
low-$T$ Nernst coefficient, i.e., we shall use Eqs. \eqref{boltzmann} to calculate
$\nu_{xy}/T$ and $\nu_{yx}/T$.


{\it Analytical considerations.}
On general grounds, the Hall conductivities obey $\sigma_{xy} = -\sigma_{yx}$ independent of the
crystal symmetry.\cite{hallnote} Such a relation does not hold for $\alpha_{xy,yx}$ in
general, however, in the low-$T$ limit the Mott relation can be derived from Boltzmann
theory,\cite{ashcroft} $\hat\alpha \propto d\hat\sigma/d\mu$, implying that
$\alpha_{xy} = -\alpha_{yx}$.
Any anisotropy in the Nernst signal is therefore arising from $\sigma_{xx} - \sigma_{yy} \neq 0$
and $\alpha_{xx} - \alpha_{yy} \neq 0$ -- both will have a piece which is linear in $\e$.
In general, the Nernst anisotropy at low $T$ will depend on the anisotropies
of the Fermi surface, of the Fermi velocities, and of the scattering rate
(the latter anisotropy is not taken into account in our calculation).

It is worth noting one limiting result: For the simplest anisotropic dispersion,
$\varepsilon_{ \bf k} = (k_x^2/m_x + k_y^2/m_y)/2$ with $m_{y,x}= (1\pm\epsilon/2)m$,
the anisotropies of $\sigma$ and $\alpha$ obtained in the Boltzmann framework \eqref{boltzmann}
{\em cancel} in the Nernst signal, i.e., the Nernst signal remains zero due to Sondheimer
cancellation.\cite{sondheimer48}
($\nu=0$ is an artifact of an energy-independent $\tau_0$.)
However, we shall obtain a large anisotropic Nernst response for non-parabolic dispersion.

\begin{figure}[!t]
\epsfxsize=2.8in
\centerline{\epsffile{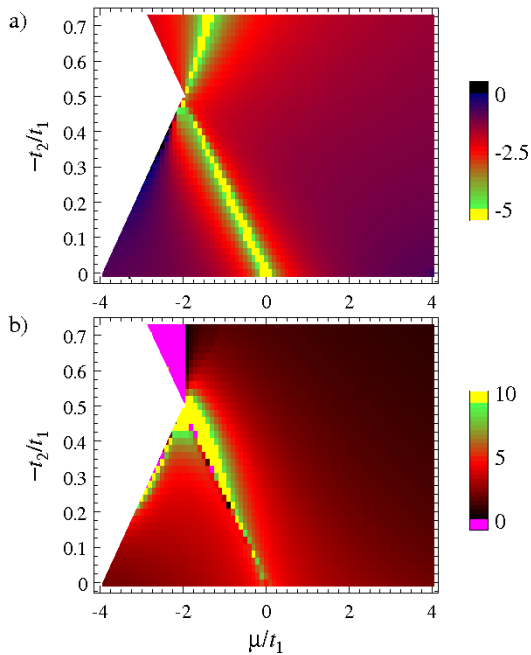}}
\caption{
Evolution of the Nernst effect as function of chemical potential $\mu$ and
second-neighbor hopping $(-t_2/t_1)$.
a) Nernst signal $\nu/T$ in the isotropic case, in
units of $2.45 \times 10^{3} {\rm V} / ({\rm K}^2 {\rm T}) \times \tau_0/{\rm s}$
for $t_1 = 1$\,eV.
b) Relative linear change of the Nernst signal with hopping anisotropy,
$d(\nu_{yx}-\nu_{xy})/(\nu d\epsilon)$.
In both panels, the white area corresponds to an empty band.
For $-t_2/t_1 > 1/2$ the band minimum moves to $(0,\pi)$, $(\pi,0)$.
For $t_2=\mu/4$ the van-Hove singularity is located at the
Fermi level, where both the Nernst signal and its sensitivity to anisotropy are maximum.
(Very close to this line the data are inaccurate due to discretization errors.)
By particle-hole symmetry, the data for $t_2/t_1>0$ can be read off using
$\mu\to-\mu$.
}
\label{fig:color}
\end{figure}

{\it Numerical results.}
We start with a survey of the Nernst signal in the $C_4$-symmetric case.
$\nu/T$ is shown in Fig.~\ref{fig:color}a as function of the chemical potential and the
second-neighbor hopping $t_2$, with $t_3$ kept zero.
While $\nu/T$ is negative for all parameters, it is small for approximately circular-shaped Fermi
surfaces, including the cases of small and large band filling for $|t_2/t_1|<1/2$.
(Note that $\nu/T$ remains non-zero in the limit of zero filling, as {\em both} the
lattice effects and the Fermi energy vanish in this limit.)
In contrast, the Nernst signal is large near the line $t_2=\mu/4$ where the 2d van-Hove
singularity is at the Fermi surface.

We now consider the sensitivity of the Nernst signal to a nematic distortion, i.e., a
hopping anisotropy. Both $\nu_{xy}/T$ and $\nu_{yx}/T$ vary linearly with $\epsilon$,
thus we plot the relative linear variation $(1/\nu)d(\nu_{yx}-\nu_{xy})/d\epsilon|_{\epsilon=0}$
in Fig.~\ref{fig:color}b.
As the Nernst signal itself, its variation is small near the band edges for
$|t_2/t_1|<1/2$, while it is large near the van-Hove singularity.
Remarkably, in a large regime of parameters the relative variation is 2\ldots5,
i.e., the Nernst signal responds to anisotropies 2\ldots5 times stronger compared
to the kinetic energy.
The sign of the anisotropy, $\nu_{yx}-\nu_{xy}$,
is robust and such that the (negative) Nernst signal is enhanced if
the thermal gradient is applied along the direction of stronger hopping.

Hole-doped cuprates are located at $-t_2/t_1 \approx 0.2\ldots0.4$
and $\mu/t_1=-0.5\ldots0$, i.e., rather close to the van-Hove regime.
We now discuss in more detail the thermoelectric response for such parameters.
We choose $t_1=0.38$\,eV and present results as function of hopping anisotropy
for different (fixed) band fillings $n$ and different $t_{2,3}$.
In \ybco, the bare hopping anisotropy extracted from band-structure calculations
is $|\e| \approx 3\ldots4\%$ with $t_a<t_b$ (i.e. $\epsilon<0$ for $\hat x$
along the crystalline a axis) --
this effect is primarily due to the coupling between planes and chains.\cite{andersen08}
Nematic order is expected to strongly amplify this anisotropy, and theoretical treatments
have suggested anisotropies up to $|\e| = 20\%$.\cite{yamase06}


\begin{figure}[!t]
\epsfxsize=2.6in
\centerline{\epsffile{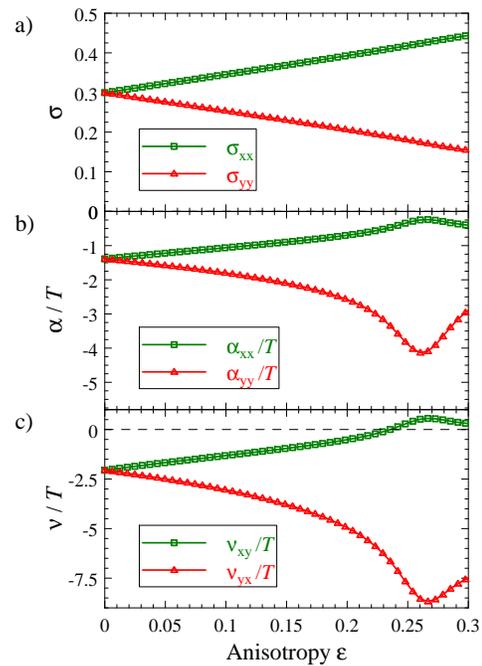}}
\caption{
Anisotropic response functions for nematic order, with
hopping anisotropy $\epsilon$. The hopping parameters are
$t_1=0.38$\,eV,
$t_2=-0.4 t_1$,
$t_3=0$, the band filling is fixed to $n=0.86$.
a) Longitudinal conductivity $\sigma_{xx}$, $\sigma_{yy}$ and
b) thermoelectric tensor $\alpha_{xx}/T$, $\alpha_{yy}/T$;
c) Nernst coefficient $\nu_{xy}/T$, $\nu_{yx}/T$.
The units are
$\sigma$ $[3.19 \times 10^{20} / (\Omega {\rm m}) \times \tau_0/{\rm s}]$ (using the c-axis lattice constant of \ybco),
$\alpha/T$ $[2.36 \times 10^{12} {\rm V} / ({\rm K}^2 \Omega {\rm m})  \times \tau_0/{\rm s}]$,
$\nu/T$ as in Fig.~\ref{fig:color}.
}
\label{fig:contrib}
\end{figure}

Fig.~\ref{fig:contrib} shows the anisotropic elements of the response tensors
$\hat\sigma$ and $\hat\alpha$, together with the Nernst coefficient,
for $t_2/t_1= -0.4$, $t_3=0$ and band filling $n=0.86$.
While the anisotropy in the longitudinal conductivity $\sigma_{xx,yy}$ is moderate,
Fig.~\ref{fig:contrib}a,
there is a substantial anisotropy in $\alpha_{xx,yy}$,
Fig.~\ref{fig:contrib}b.
Remarkably, both anisotropies constructively interfere to generate a large Nernst
anisotropy, Fig.~\ref{fig:contrib}c.
The signal peaks at some finite $\epsilon$ where the Fermi-surface topology changes as a
function of the anisotropy -- this point again corresponds to a (now anisotropic) van-Hove
singularity.
Note that $\nu_{xy}$ can even change sign at some finite $\epsilon$,
leading to a formal divergence of $\nu_{yx}/\nu_{xy}$.
(Other additive contributions to $\nu$, e.g., from pairing fluctuations,
may render this statement insignificant.)

A few remarks are in order.
First, the thermopower $S=\alpha_{xx}/\sigma_{xx}$ comes out to be negative,
Fig.~\ref{fig:contrib}b, also for $\epsilon=0$.
This is in contrast to the common wisdom that a single hole-like Fermi surface leads to a
positive thermopower. Indeed, we obtain a positive (negative) thermopower in the limits
of large (small) band filling, but the sign change in the $\mu$--$t_2$ plane is {\em not}
tied to the topological change of the Fermi surface (i.e. the van-Hove singularity), but
instead occurs above half filling for $t_2<0$.
Second, the sign of the anisotropy in $\sigma$, Fig.~\ref{fig:contrib}a, is robust.
This is physically transparent: $t_x>t_y$ implies a larger velocity along $\hat x$, and
hence generically $\sigma_{xx}>\sigma_{yy}$.
Third, the sign of the anisotropy $(\alpha_{xx}-\alpha_{yy})$, Fig.~\ref{fig:contrib}b, changes in the
$\mu$--$t_2$ plane exactly at the van-Hove singularity.
Notably, in Eq.~\eqref{nernstsignal} this sign change is compensated by the
similar sign change of the Hall conductivity $\sigma_{xy}$, such that sign of the Nernst
anisotropy is robust.

\begin{figure}[!t]
\epsfxsize=3.5in
\centerline{\epsffile{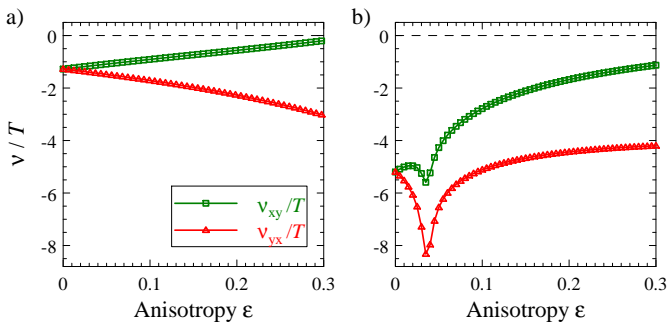}}
\caption{
Nernst coefficient $\nu/T$ as function of the hopping
anisotropy $\epsilon$, for parameters $t_1=0.38$\,eV and
a) $t_2=-0.32 t_1$, $t_3= -0.5 t_2$, $n=0.875$, and
b) $t_2=-0.2 t_1$, $t_3=0$, $n=0.875$.
The units are as in Fig.~\ref{fig:color}.
}
\label{fig:full}
\end{figure}

The influence of model parameters is illustrated in Fig.~\ref{fig:full}.
For finite $t_3/t_2<0$, the van-Hove singularity is outside the regime of $\epsilon$ and
$n$ which is relevant for cuprates. (This conclusion may be changed by bilayer splitting,
see below and Ref.~\onlinecite{yamase06}.)
Fig.~\ref{fig:full} also makes clear that the response of $\nu/T$ to moderate anisotropies
is strongly non-linear near a van-Hove point.

{\it Discussion.}
The large Nernst anisotropy for a nematic state prompts the question how
this compares to other states with rotation symmetry breaking. For cuprates, the prime
candidate is stripe order.
Following Ref.~\onlinecite{hackl09b}, we have also calculated the Nernst anisotropy in various
stripe states with real-space period 8 and 16. In most cases, the Nernst anisotropy for
realistic parameter values is moderate, i.e., less than a factor of two. Exceptions are
states with extremely elongated Fermi pockets, which we only encountered for period-4
charge-only stripes (and for which no evidence exists in \ybco).
Hence, stripe order is unlikely to explain the large Nernst anisotropy in \ybco.\cite{taill10b}

We thus confirm the interpretation of the data in Ref.~\onlinecite{taill10b},
attributing the large anisotropy for $80\,{\rm K}<T<150\,{\rm K}$ to a nematic state
(while stripe-like order may set in at lower temperatures).
Indeed, using $\epsilon=-0.2$ from Ref.~\onlinecite{yamase06},
we can semi-quantitatively match the experimental results for \ybco:
We obtain a resistivity anisotropy
$\sigma_b/\sigma_a \equiv \sigma_{yy}/\sigma_{xx} \approx 1.5 \ldots 2.5$
(Ref.~\onlinecite{ando02}) and a Nernst anisotropy
$\nu_b/\nu_a \equiv \nu_{xy}/\nu_{yx} \approx 4\ldots10$ (Ref.~\onlinecite{taill10b}).

Let us briefly discuss effects which were neglected in our calculation.
A vertical dispersion from inter-layer coupling will smear the 2d van-Hove singularity,
i.e., will remove the sharp peak in the Nernst signal as function of filling or distortion.
However, as the Nernst signal is enhanced and very sensitive to anisotropy over a large
range of parameters, this will not qualitatively change our conclusions.
For \ybco, the main effect of inter-layer coupling is a bilayer splitting of the
dispersion, which will move one of the van-Hove singularities closer to the physical
parameter regime, likely enhancing the Nernst signal.
Finally, at elevated temperatures, the scattering rate will be both anisotropic and
energy-dependent. While it is known for cuprates that the antinodal scattering rates become
much larger than the nodal ones, our conclusions would only be modified if the scattering
near the two inequivalent {\em antinodes} becomes different by a factor of order 5 --
this we consider unlikely.


{\it Conclusions.}
Our key result is a surprisingly large sensitivity of the Nernst anisotropy to
symmetry-breaking Fermi-surface distortions, rendering the Nernst effect a unique tool
to detect electron-nematic order.
The effect is particularly strong for non-parabolic band dispersion, as occurs near
van-Hove singularities and/or half-filling.
We have provided a concrete theory for the measured Nernst anisotropies
in \ybco,\cite{taill10a,taill10b} pointing towards electron-nematic order in the
pseudogap regime.
We propose to search for Nernst anisotropies in other candidate compounds
for such order, such as \srruo. Our results in Fig.~\ref{fig:contrib} suggest
that anisotropies of the thermopower are worth investigating as well.



We are particularly grateful to L. Taillefer for sharing data prior to publication.
We thank S. Kivelson, A. J. Millis, S. Sachdev, and L. Taillefer for discussions,
and S. Sachdev for collaborations on related work.
This research was supported by the DFG through
SFB 608 (K\"oln) and the Research Unit FG 538.



\begin{thebibliography}{}
\bibitem{directfoot}
A probe which directly couples to the bulk quadrupolar order parameter of an
electron-nematic phase is usually not available.
\bibitem{signnote}
For fixed magnetic field $\vec B || \hat z$,
a $C_4$-symmetric situation yields $\vartheta_{xy} = -\vartheta_{yx}$ from
Eqs.~(\ref{thermoelectrics},\ref{nernstsignal}).
Here we adopt the convention (also used experimentally)
that the three vectors $\vec E$, $\vec\nabla T$, and $\vec B$ form a right-handed
system for the measurements of both $\nu_{xy}$ and $\nu_{yx}$,
which may be achieved by switching the sign of the magnetic field.
Then, $\nu_{xy} = \nu_{yx}$ for $C_4$ symmetry.
\bibitem{hallnote}
This follows from the Kubo expressions for $\sigma_{xy,yx}$
or, for low $T$, from the geometric interpretation of the Hall conductivity, see:
N. P. Ong, \prb {\bf 43}, 193 (1991).
\end{thebibliography}
\end{document}